\documentclass[aps,prstab,preprint,groupedaddress]{revtex4}
\usepackage{amsmath}
\usepackage{graphicx}
\usepackage{color}

\begin{document}


\title{Analytic fluid theory of beam spiraling in high-intensity cyclotrons}


\author{A.J. Cerfon} 
\email[]{cerfon@cims.nyu.edu}
\affiliation{Courant Institute of Mathematical Sciences, New York University, 251 Mercer St, New York NY 10012, United States}
\author{J.P. Freidberg, F.I. Parra}
\affiliation{MIT Plasma Science and Fusion Center, 167 Albany St, Cambridge MA 02139, USA}

\author{T.A. Antaya}
\affiliation{Ionetix Corporation}

\date{\today}

\begin{abstract}
Using a two-dimensional fluid description, we investigate the nonlinear radial-longitudinal dynamics of intense beams in storage rings and cyclotrons. With a multiscale analysis separating the time scale associated with the betatron motion and the slower time scale associated with space-charge effects, we show that the longitudinal-radial vortex motion can be understood in the frame moving with the charged beam as the nonlinear advection of the beam by the $\mathbf{E}\times\mathbf{B}$ velocity field, where $\mathbf{E}$ is the electric field due to the space charge and $\mathbf{B}$ is the external magnetic field. This interpretation provides simple explanations for the stability of round beams and for the development of spiral halos in elongated beams. By numerically solving the nonlinear advection equation for the beam density, we find that it is also in quantitative agreement with results obtained in PIC simulations.
\end{abstract}


\maketitle

\section{Introduction}

In recent years, novel uses of particle accelerators for nuclear energy, nuclear security, medicine and material science applications have triggered the design, development and construction of moderate size, high intensity cyclotrons. The higher beam intensities expected in these new machines provide more powerful tools for these new endeavours, but also come with very stringent uncontrolled beam loss requirements. Satisfying these low beam loss criteria requires detailed knowledge of the beam dynamics in such machines. In particular, collective modes and instabilities associated with space charge effects have to be taken into account and need to be studied and understood. Theory and simulation tools are being developed to that purpose. The computational effort is crucial because of the complexity of the problem at hand. One can hardly imagine a purely analytic study that would accurately account for the complex magnetic geometry of modern cyclotrons and at the same time properly model the nonlinear beam dynamics due to space charge effects. However, analytical theory work is equally as important in order to provide an interpretation for the results from complicated, all-inclusive simulations as well as to identify the basic contributing mechanisms.

Interestingly, the computational effort in high-intensity cyclotrons is almost entirely focused on Particle-In-Cell (PIC) methods \cite{Hockney_Eastwood},\cite{Cousineau},\cite{YangAdel},\cite{Adel}. PIC methods have a number of advantages. They are intuitive, naturally lead to parallel solvers, and can easily take exact magnetic geometries as inputs. Nowadays, PIC methods are commonly used to design, interpret and corroborate experiments in which space charge effects play a significant role in the beam dynamics \cite{Holmes},\cite{Zhang},\cite{Pozdeyev1}. However, precisely because of their conceptual simplicity, PIC codes tend not to yield as much insights into the physical mechanisms at work in a given observed phenomena as do continuum descriptions. Theoretical work based on continuum descriptions, whether kinetic or fluid, are often better suited to identify the key nondimensional parameters and to derive the typical scalings for the dependence of the quantities of interest on these nondimensional parameters. Analytic and numerical work based on continuum descriptions are therefore a necessary and useful complement to PIC simulations

In this article, we illustrate the constructive interplay of experiments, PIC simulations and theory in the study of a topic of high interest in beam dynamics in isochronous cyclotrons, namely the coherent longitudinal-radial vortex motion associated with the nonlinear interplay of radial and longitudinal space-charge forces. It has been oberved in PIC simulations \cite{YangAdel},\cite{Adam},\cite{Koscielniak},\cite{Bertrand} that a coasting beam which is elongated in the longitudinal direction (the direction of propagation of the beam) develops a spiral galaxy shape due to space-charge effects. This phenomenon, which has sometimes been called the ``spiraling instability" \cite{Adam}, can play a positive role when the spiral arms are lost after a few turns, at low beam energy, and the center of the spiral becomes a very compact stable beam \cite{YangAdel},\cite{Bertrand},\cite{AdamAdel},\cite{Seidel}. The spiraling of the beam can also play a negative role. It is thought to lead to beam breakup, as confirmed by experiment and PIC simulations \cite{Pozdeyev2},\cite{Bi}. At lower beam currents, the spiral arms can also grow over longer time scales, and therefore form a beam halo only after a large number of turns, at which point it has been accelerated to high energies. 

Until now, aside from PIC simulations, the theoretical explanations of beam spiraling have been based on single-particle pictures \cite{Bertrand},\cite{Pozdeyev2},\cite{Bi},\cite{Chasman}. In these single-particle descriptions, the evolution of the beam in the presence of space-charge forces is extrapolated from the modifications of the motion of single particles due to this force. In contrast, using a fluid model of the charged beam we present a new self-consistent nonlinear description of the coherent radial-longitudinal vortex motion and give an intuitive explanation for the formation of the spiral galaxy shape in an elongated charged particle beam. With a multiscale analysis that separates the time scale associated with the betatron motion and the time scale associated with the space-charge force, we show that in the frame rotating with the center of the beam, the evolution of the beam on the space-charge time scale is determined by the nonlinear advection of the beam density by the $\mathbf{E}\times\mathbf{B}$ velocity. Here, $\mathbf{E}$ is the self electric field, and $\mathbf{B}$ is the external cyclotron magnetic field. With this new fluid picture of the radial-longitudinal vortex motion, the observed behavior in experiments and PIC simulations is easily understood, and beam spiraling has a simple explanation. 

Consider first a round beam, whose charge density is only a function of its radius. In the frame rotating with the beam, the self electric field of such a charge distribution is only along the direction of the beam radius, and the $\mathbf{E}\times\mathbf{B}$ velocity vector in the radial-longitudinal plane is therefore purely perpendicular to the beam radius at every point. This implies that the advection in the $\mathbf{E}\times\mathbf{B}$ field will have no effect on the radial distribution of the beam density: a round beam thus has a stationary distribution. In an elongated beam however, the cylindrical symmetry of the problem is broken since the density distribution depends both on radius and angle. The beam density distribution will thus be distorted by the $\mathbf{E}\times\mathbf{B}$ velocity field. Since the $\mathbf{E}\times\mathbf{B}$ velocity is largest at the extremities of the beam, the distortion will be largest away from the beam center, which leads to the formation of spiral galactic arms. This intuitive explanation is confirmed by simulations in which we numerically solve the nonlinear fluid equations, and we demonstrate quantitative agreement between our fluid description and PIC simulations.  

The structure of the paper is as follows. In Section \ref{sec:two}, we describe the geometry of the problem and introduce the fluid description of the beam which we use for our analysis. In Section \ref{sec:three}, we carry out a multiscale expansion of the fluid equations, and derive the advection equation for the evolution of the beam density on the time scale associated with the space charge force. We solve this equation numerically in Section \ref{sec:four} for different cases of interest, and demonstrate both the formation of spiral galactic arms for elongated beams and the stability of round beams. We end with conclusions in Section V. 

\section{Fluid description of the beam}\label{sec:two}

One of the main motivations for this work is to offer an analytic derivation of the self-consistent beam vortex motion under the effect of space charge. Such a goal implies that we will have to make simplifications in order to make our analysis tractable. In making these simplifications, our guiding principle is to simplify the geometric aspects of the problem enough so that only a minimal amount of beam physics assumptions need to be made.

\subsection{Geometry of the problem}

It has been observed in PIC simulations that the details of the beam description along the vertical direction, i.e. along the direction of the main applied magnetic field, have a very limited influence on the longitudinal-radial vortex motion \cite{YangAdel},\cite{Koscielniak},\cite{Adam2}. To make our analysis more tractable, we therefore restrict our study to the two-dimensional problem of a nonrelativistic coasting beam in the radial-longitudinal plane, immersed in a homogeneous magnetic field along the $z$ axis, which coincides with the vertical direction. Throughout our analysis we will thus have $\mathbf{B}=B\mathbf{e}_{z}$ with $B=constant$, and $\partial/\partial z=0$ will hold. One implication of these assumptions is that the formation of spiral arms does not depend on relativistic effects or field gradient along the vertical direction.
 
\begin{figure}[ht]
\centering
\includegraphics[width=\textwidth]{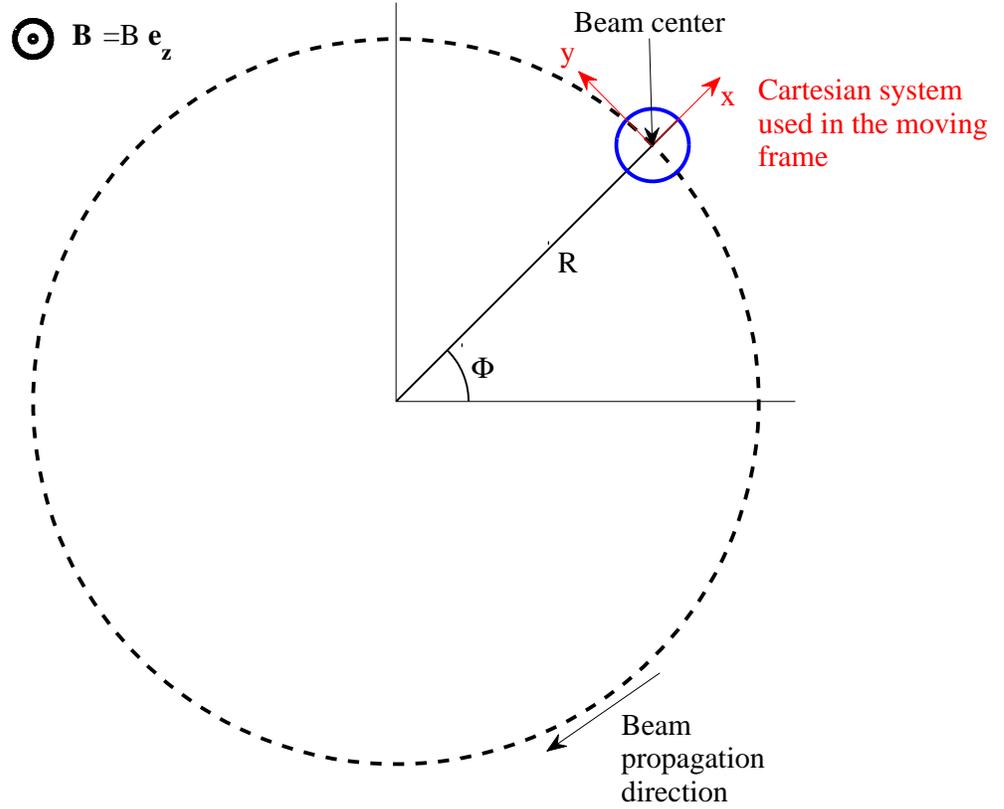}
\caption{Polar coordinates $(R,\Phi)$ used to define the change of reference frame to the frame rotating with the beam, and (in red) Cartesian coordinate system used in the rotating frame}
\label{fig:lab_frame}
\end{figure}

In the radial-longitudinal plane, i.e. the plane perpendicular to the magnetic field, it is most convenient to study space-charge effects in the frame moving with the coasting beam. The transformation from the laboratory frame to the moving frame is best understood in terms of the polar coordinates $(R,\Phi)$ associated with the cyclotron geometry and shown in Figure \ref{fig:lab_frame}. 

\begin{figure}[ht]
\centering
\includegraphics[width=\textwidth]{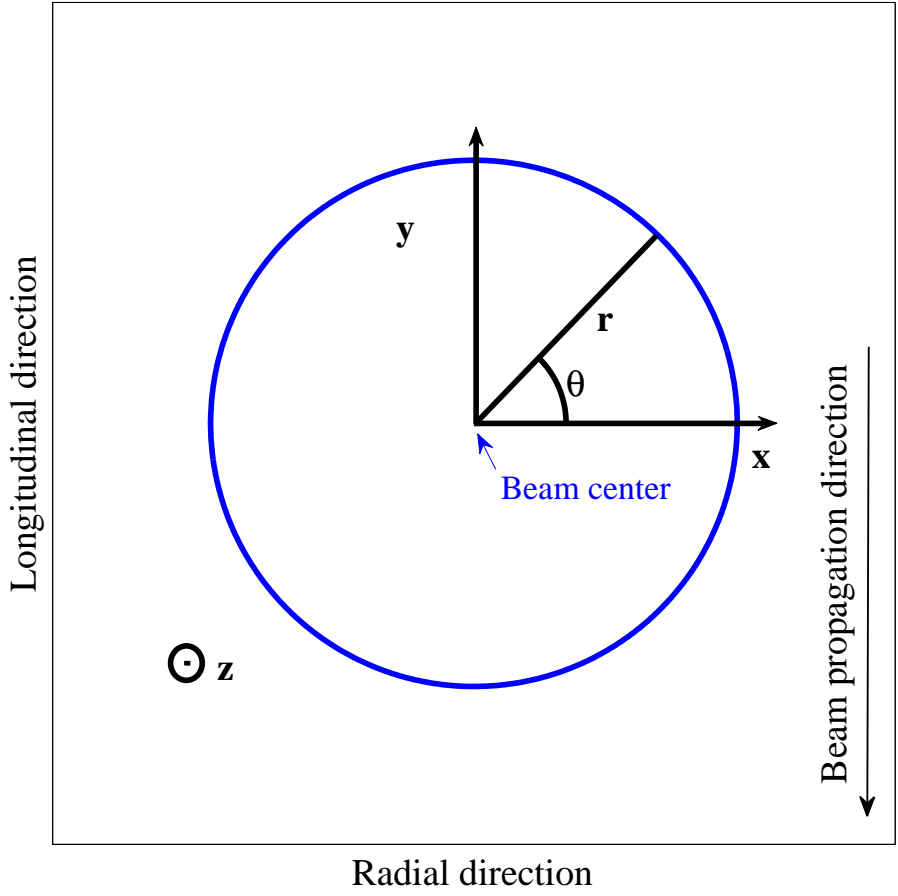}
\caption{$(x,y)$ and $(r,\theta)$ coordinate systems used for the numerical simulations and the physical interpretation of results.}
\label{fig:coordinate_pic}
\end{figure}

We consider a single-species intense ion beam, and for the simplicity of the notation, the ions are singly charged: $q=e$. The ion mass is $m$, so that the cyclotron frequency is $\omega_{c}=eB/m$. The change of reference frame to the frame rotating with the beam is given by the transformation
\begin{equation}
R'=R\qquad\Phi'=\Phi+\omega_{c}t\qquad t'=t\qquad v_{R}^{'}=v_{R}\qquad v_{\Phi}^{'}=v_{\Phi}+R\omega_{c}
\label{eq:change_frame}
\end{equation}
where the primed quantities represent quantities in the moving frame. Using the relations in Eq.\eqref{eq:change_frame}, we easily find
\begin{equation}
\frac{\partial}{\partial R}=\frac{\partial}{\partial R'}\qquad\frac{\partial}{\partial \Phi}=\frac{\partial}{\partial\Phi'}\qquad \frac{\partial}{\partial t}=\frac{\partial}{\partial t'}+\omega_{c}\frac{\partial}{\partial\Phi'}
\label{eq:change_frame_partial}
\end{equation}
In the next section, we will use these relations to express the fluid equations in the frame moving with the beam. Before doing so, we briefly discuss the two coordinate systems we will use in the local frame moving with the beam, which are shown in Figure \ref{fig:coordinate_pic}. The first system is a two-dimensional Cartesian coordinate system $(x,y)$ with the origin at the beam center, and defined such that $(x,y,z)$ is a right-handed orthogonal coordinate system with the $x$-axis coinciding with the radial direction, i.e the main radius of the cyclotron, and the $y$-axis coinciding with the longitudinal direction. This is the coordinate system also plotted in red in Figure \ref{fig:lab_frame}. In this coordinate system, the ion beam travels in the $-y$ direction. The second system is a polar coordinate system $(r,\theta)$ centered at the center of the beam, defined such that $(r,\theta,z)$ is a right-handed orthogonal cylindrical coordinate system and such that $x=r\mbox{cos}\theta$ and $y=r\mbox{sin}\theta$. It is equipped with the orthonormal basis $(\mathbf{e}_{r},\mathbf{e}_{\theta})$, where the unit vectors $\mathbf{e}_{r}$ and $\mathbf{e}_{\theta}$ are defined in terms of the Cartesian unit vectors $\mathbf{e}_{x}$ and $\mathbf{e}_{y}$ according to $\mathbf{e}_{r}=\mbox{cos}\theta\mathbf{e}_{x}+\mbox{sin}\theta\mathbf{e}_{y}$ and $\mathbf{e}_{\theta}=-\mbox{sin}\theta\mathbf{e}_{x}+\mbox{cos}\theta\mathbf{e}_{y}$. The Cartesian coordinate system will be mostly used for the numerical simulations in section \ref{sec:four}, and the polar coordinate system is the most appropriate for the physical interpretation of results.

\subsection{Fluid description of the beam}

At the beam temperatures and densities relevant to high intensity cyclotrons, the charged particle beam is a nonneutral plasma that can be considered collisionless \cite{YangAdel}. Then, by taking the first two moments of the collisionless Vlasov equation, we obtain the following exact fluid equations involving the beam density $n$, velocity $\mathbf{v}$ and pressure tensor $\mathbf{P}$ \cite{Davidson}:
\begin{equation}
\frac{\partial n}{\partial t}+\nabla\cdot\left(n\mathbf{v}\right)=0\label{eq:moment_density}
\end{equation}
\begin{equation}
mn\left(\frac{\partial\mathbf{v}}{\partial t}+\mathbf{v}\cdot\mathbf{\nabla}\mathbf{v}\right)+\mathbf{\nabla}\cdot\mathbf{P}=en\left(\mathbf{E}+\mathbf{v}\times{\mathbf{B}}\right)\label{eq:moment_momentum}
\end{equation}
In Eq.\eqref{eq:moment_momentum}, $\mathbf{E}$ is the self electric field, which is determined by solving 
\begin{equation}
\nabla\cdot\mathbf{E}=\frac{en}{\epsilon_{0}}\label{eq:divE}
\end{equation}
$\mathbf{B}$ is the external magnetic field $\mathbf{B}=B\mathbf{e}_{z}$ since the self magnetic field is negligible in the nonrelativistic limit \cite{Lund}. Using the relations in Eq. \eqref{eq:change_frame_partial}, we can rewrite Equations \eqref{eq:moment_density}, \eqref{eq:moment_momentum} and \eqref{eq:divE} in the frame rotating with the beam. We find

\begin{equation}
\frac{\partial n'}{\partial t'}+\nabla'\cdot\left(n'\mathbf{v}'\right)=0\label{eq:moment_density_rotating}
\end{equation}
\begin{equation}
mn'\left(\frac{\partial\mathbf{v}'}{\partial t'}+\mathbf{v}'\cdot\mathbf{\nabla}'\mathbf{v}'\right)+\mathbf{\nabla}'\cdot\mathbf{P}'=en'\left(\mathbf{E}'-\mathbf{v}'\times{\mathbf{B}}\right)\label{eq:moment_momentum_rotating}
\end{equation}
\begin{equation}
\nabla'\cdot\mathbf{E}'=\frac{en'}{\epsilon_{0}}\label{eq:divE_rotating}
\end{equation}
where as before the primed quantities represent quantities in the moving frame. From now on we will always work in the frame rotating with the beam, so to simplify the notation we will drop the prime symbols in the remainder of the article. Note that in the moving frame, there is a minus in the magnetic field force. It is due to the Coriolis force. 

In the frame rotating with the beam the electrostatic approximation holds \cite{YangAdel}, so we write $\mathbf{E}=-\nabla\phi$ where $\phi$ is the electrostatic potential. Then, if we call $a$ the characteristic size of the beam (its radius if the beam is circular), $N_{0}$ its peak density and $T_{0}$ its peak temperature, renormalizing the fluid quantities according to
\begin{eqnarray*}
t&\rightarrow&\omega_{c}t\qquad\nabla \rightarrow a\nabla\qquad\mathbf{v}\rightarrow\frac{\mathbf{v}}{a\omega_{c}}\\
n&\rightarrow&\frac{n}{N_{0}}\qquad \mathbf{P}\rightarrow\frac{\mathbf{P}}{N_{0}T_{0}}\;\;\;\;\phi\rightarrow\frac{\epsilon_{0}}{eN_{0}a^{2}}\phi
\end{eqnarray*}
leads to the following set of nondimensional fluid equations in the frame rotating with the beam:
\begin{eqnarray}
\frac{dn}{dt}&+&n\nabla\cdot\mathbf{v}=0\label{eq:continuity_moving_intermed}\\
\frac{d\mathbf{v}}{dt}&+&\mathbf{v}\times\mathbf{e}_{z}=-\delta^{2}\left(\nabla\phi+\frac{\alpha^{2}}{n}\nabla\cdot \mathbf{P}\right)\label{eq:momentum_moving_intermed}\\
\nabla^{2}\phi&=&-n\label{eq:Poisson_intermed}
\end{eqnarray}
Here, $d/dt\equiv\partial/\partial t+\mathbf{v}\cdot\nabla$ and $\delta^{2}=\omega_{p}^{2}/\omega_{c}^{2}$ where $\omega_{p}$ is the plasma frequency, $\omega_{p}^{2}=N_{0}e^{2}/m\epsilon_{0}$. The parameter $\alpha$ is defined by $\alpha^{2}=T_{0}/ma^{2}\omega_{p}^{2}=\lambda_{D}^{2}/a^{2}$ where $\lambda_{D}$ is the Debye length. The experimental regime of interest for high intensity beams is $\alpha<1$, so that space-charge effects dominate over temperature effects.

It is clear from Eqs.\eqref{eq:continuity_moving_intermed}-\eqref{eq:Poisson_intermed} that the parameter determining the importance of space-charge effects on the behavior of the beam is $\delta^{2}$, the ratio of the beam plasma frequency squared to the cyclotron frequency squared. This was observed in a slightly different context in Eq.(12) in \cite{Pozdeyev2} and in \cite{Reiser}. In particular, in \cite{Reiser} the key role of the parameter $\chi$ is emphasized, with $\chi$ defined so that $\chi=\delta^{2}/2$ in our notation. As noted in \cite{Reiser}, all cyclotrons and rings satisfy $\delta^{2}\leq1$, and most machines satisfy $\delta^{2}\ll1$. The parameter $\delta^{2}$ can then be used to derive an approximate form for the pressure tensor $\mathbf{P}$ in the limit of small $\delta$, as we will do in the next paragraph. It can also be used to perform a multiscale analysis of the resulting beam fluid equations which separates the fast motion associated with betatron oscillations from the slower motion due to space-charge and thermal forces. This is precisely what we do in Section \ref{sec:three}.

Equations \eqref{eq:continuity_moving_intermed}-\eqref{eq:Poisson_intermed} cannot be solved as such since we are still facing the problem of closure of the hierarchy of moment equations \cite{Davidson}. However, a key point of our analysis is that with our simplified geometry, a single assumption we make concerning the beam physics is sufficient to address the closure issue. Specifically, we assume that the amplitude of the betatron oscillations in the frame moving with the beam is small compared to the characteristic size of the beam by the ratio $\delta$, as we will discuss in more detail in Section \ref{sec:three} when we describe the multiscale expansion. In this regime, we prove in the Appendix, starting from the collisionless Vlasov equation, that the pressure tensor $\mathbf{P}$ must be gyrotropic to lowest order:
\begin{equation}\label{eq:gyrotropic}
\mathbf{P}=p_{\perp}\mathbf{I}+\left(p_{\parallel}-p_{\perp}\right)\mathbf{b}\mathbf{b}+O(\delta)
\end{equation}
where $\mathbf{b}=\mathbf{B}/B$ is equal to $\mathbf{e}_{z}$ in our geometry, $p_{\perp}$ is the pressure in the radial-longitudinal plane, and $p_{\parallel}$ the pressure along the direction of the magnetic field. The divergence of the gyrotropic pressure tensor is easily calculated:
\begin{equation}
\nabla\cdot\mathbf{P}=\nabla p_{\perp}+\left(p_{\parallel}-p_{\perp}\right)\mathbf{b}\cdot\nabla\mathbf{b}+\mathbf{b}\cdot\nabla\left(p_{\parallel}-p_{\perp}\right)\mathbf{b}+\left(p_{\parallel}-p_{\perp}\right)\nabla\cdot\mathbf{b}\mathbf{b}+O(\delta)
\end{equation}
Now, since $\mathbf{b}\cdot\nabla\mathbf{b}=\mathbf{e}_{z}\cdot\nabla\mathbf{e}_{z}=0$, in the radial-longitudinal plane the divergence of the pressure tensor takes the form of a pressure gradient to lowest order in $\delta$:
\begin{equation}\label{eq:pressure_grad_zero}
\left(\nabla\cdot\mathbf{P}\right)_{\perp}=\nabla p_{\perp}+O(\delta)
\end{equation}
In the multiscale analysis presented in Section \ref{sec:three}, we will need to expand the beam fluid equations to second order in $\delta$. Since in Eq.\eqref{eq:momentum_moving_intermed} the pressure term is  multiplied by $\delta^{2}$, we only need an expression for the pressure tensor to zeroth order in $\delta$. This is precisely what we have in Eq. \eqref{eq:pressure_grad_zero}. In the rotating frame, the appropriate fluid equations for the beam thus take the following form
\begin{eqnarray}
\frac{dn}{dt}&+&n\nabla\cdot\mathbf{v}=0\label{eq:continuity_moving}\\
\frac{d\mathbf{v}}{dt}&+&\mathbf{v}\times\mathbf{e}_{z}=-\delta^{2}\left(\nabla\phi+\frac{\alpha^{2}}{n}\nabla p\right)\label{eq:momentum_moving}\\
\nabla^{2}\phi&=&-n\label{eq:Poisson}
\end{eqnarray}
where for the simplicity of the notation we have dropped the $\perp$ symbol for $p$ since we only consider the two-dimensional radial-longitudinal plane and $p_{\parallel}$ never enters in the analysis.

Equations \eqref{eq:continuity_moving}-\eqref{eq:Poisson} are the fluid equations we solve in the remainder of this article, using a multiscale analysis. Note that these equations are not fully closed in the sense that an equation for the evolution of the pressure $p$ is missing. As we will show shortly, such an equation is not necessary. Indeed, when the pressure term can be expressed as an exact gradient in the momentum equation, pressure effects do not enter in the final equation for the evolution of the beam density on the space charge time scale. 

\section{Multiscale analysis of the fluid equations}\label{sec:three}
\subsection{Multiscale expansion}
Given the difference in the time scale associated with the betatron oscillations and that associated with the expansion of the beam due to electrostatic and thermal forces, we perform a multiple time scale analysis of Equations \eqref{eq:continuity_moving}-\eqref{eq:Poisson}. Specifically, each quantity $Q$ is assumed to vary according to the different time scales as follows:
\begin{equation}
Q(\mathbf{r},t)=Q(\mathbf{r},t_{0},t_{2},t_{4},\ldots)=Q(\mathbf{r},t,\delta^{2}t,\delta^{4}t,\ldots)
\label{eq:separate_scales}
\end{equation}
With this formal expansion, we have
\begin{equation}
\frac{\partial Q}{\partial t}=\frac{\partial Q}{\partial t_{0}}+\delta^{2}\frac{\partial Q}{\partial t_{2}}+\ldots
\end{equation}

It is convenient for the rest of the calculation to separate the quantities $Q$ into the sum of a rapidly oscillating part $\tilde{Q}$ due to the betatron oscillations, and a slow monotonic evolution $\bar{Q}$ due space charge and thermal effects:
\begin{equation}
Q(\mathbf{r},t_{0},t_{2},\ldots)=\tilde{Q}(\mathbf{r},t_{0},t_{2},\ldots)+\bar{Q}(\mathbf{r},t_{2},\ldots)
\label{eq:separation}
\end{equation}
Observe that, by construction, $\bar{Q}$ does not depend on $t_{0}$. We will show shortly that $\tilde{Q}$ is periodic in $t_{0}$.

In the frame moving with the beam, the first non-vanishing contribution to the velocity is due to the betatron oscillations. As mentioned previously, we assume in our analysis that the amplitude of these oscillations is of order $\delta a$. Since the betatron time scale is $\omega_{c}$, this means that the first non-vanishing contribution to the velocity is of order $\delta a\omega_{c}$. In other words, the velocity comes in first order in $\delta$ in our ordering and the appropriate expansion for the relevant physical quantities is the following:
\begin{eqnarray}
n&=&\bar{n_{0}}+\delta\left(\tilde{n_{1}}+\bar{n_{1}}\right)+\delta^{2}\left(\tilde{n_{2}}+\bar{n_{2}}\right)+O(\delta^{3})\\
p&=&\bar{p_{0}}+O(\delta)\\
\phi&=&\bar{\phi_{0}}+O(\delta)\label{eq:expansion}\\
\mathbf{v}&=&\delta\tilde{\mathbf{v}}_{1}+\delta^{2}\left(\tilde{\mathbf{v}}_{2}+\bar{\mathbf{v}}_{2}\right)+O(\delta^{3})
\end{eqnarray}

We introduce this expansion in the set of equations given by Eqs.\eqref{eq:continuity_moving}-\eqref{eq:Poisson}, and solve order by order in $\delta$. We start with Poisson's equation, which we only need to zeroth order
\begin{equation}
\nabla^{2}\bar{\phi_{0}}=-\bar{n_{0}}
\label{eq:Poisson_zeroth}
\end{equation}
Turning to Eq.\eqref{eq:continuity_moving}, the mass conservation equation, we see that the first non-trivial equation arises in first order in $\delta$ and is given by
\begin{equation}
\frac{\partial\tilde{n_{1}}}{\partial t_{0}}+\nabla\cdot\left(\bar{n_{0}}\tilde{\mathbf{v}}_{1}\right)=0
\label{eq:continuity_first}
\end{equation}
This equation describes the evolution of the density on the fast time scale, due to the betatron oscillations. To next order, the mass conservation equation takes the form:
\begin{equation}
\frac{\partial \tilde{n_{2}}}{\partial t_{0}}+\frac{\partial \bar{n_{0}}}{\partial t_{2}}+\nabla\cdot\left[\left(\tilde{n_{1}}+\bar{n_{1}}\right)\tilde{\mathbf{v}}_{1}+\bar{n_{0}}\left(\tilde{\mathbf{v}}_{2}+\bar{\mathbf{v}}_{2}\right)\right]=0
\label{eq:continuity_second}
\end{equation}

Averaging this equation over the fast time scale, we get, because of the periodicity of the oscillating quantities in $t_{0}$:
\begin{equation}
\frac{\partial\bar{n_{0}}}{\partial t_{2}}+\nabla\cdot\left(<\tilde{n_{1}}\tilde{\mathbf{v}}_{1}>+\bar{n_{0}}\bar{\mathbf{v}}_{2}\right)=0
\label{eq:continuity_second_averaged}
\end{equation}
where we have introduced the notation
\begin{displaymath}
<Q>=\frac{1}{2\pi}\int_{0}^{2\pi}Qdt_{0}
\end{displaymath}
Eq.\eqref{eq:continuity_second_averaged} describes the evolution of the bunch density on the slow time scale. This is where the influence of the electrostatic effects and the beam temperature will enter, and this is therefore all we need from the mass conservation equation. The purpose of the remainder of the article is to solve this equation, first by deriving expressions for $<\tilde{n_{1}}\tilde{\mathbf{v}}_{1}>$ and $\bar{\mathbf{v}}_{2}$ in terms of zeroth order quantities, and by then solving the resulting equation numerically. The necessary expressions are obtained from the momentum equation. To first order in $\delta$, we have 
\begin{equation}
\frac{\partial\tilde{\mathbf{v}}_{1}}{\partial t_{0}}+\tilde{\mathbf{v}}_{1}\times\mathbf{e}_{z}=\mathbf{0}
\label{eq:momentum_first}
\end{equation}
and by averaging the second order momentum equation over the fast time scale as was done before, we find
\begin{equation}
\bar{\mathbf{v}}_{2}=<\tilde{\mathbf{v}}_{1}\cdot\nabla\tilde{\mathbf{v}}_{1}>\times\mathbf{e}_{z}+\nabla\bar{\phi_{0}}\times\mathbf{e}_{z}+\frac{\alpha^{2}}{\bar{n_{0}}}\nabla\bar{p_{0}}\times\mathbf{e}_{z}
\label{eq:momentum_second_averaged}
\end{equation}

At this point, all the relevant equations have been derived. From Eqs.\eqref{eq:continuity_first}, \eqref{eq:momentum_first} and \eqref{eq:momentum_second_averaged} we can derive expressions for $\tilde{n}_{1}$, $\tilde{\mathbf{v}}_{1}$ and $\bar{\mathbf{v}}_{2}$ in terms of zeroth order quantities and initial conditions, which we can then insert in Eq.\eqref{eq:continuity_second_averaged} to obtain the desired equation for the evolution of the beam density on the space-charge time scale. This is done in the next subsection. 

\subsection{Solving the equations}
We start the calculation with the description of the betatron motion to lowest order, i.e. $\tilde{\mathbf{v}}_{1}$, and its effect on the bunch density $\tilde{n}_{1}$. $\tilde{\mathbf{v}}_{1}$ is given by Eq.\eqref{eq:momentum_first}, which is just the equation of a particle immersed in a uniform magnetic field. It is easily solved:
\begin{equation}
\tilde{\mathbf{v}}_{1}(\mathbf{r},t_{0},t_{2},\ldots)=\mathbf{u_{1}}\mbox{cos}t_{0}+\mathbf{e}_{z}\times\mathbf{u}_{1}\mbox{sin}t_{0}\label{eq:v_betatron_solved}
\end{equation}
where $\mathbf{u}_{1}(\mathbf{r},t_{2},\ldots)$ is the initial betatron velocity, i.e. $\tilde{\mathbf{v}}_{1}$ at time$\ t_{0}=0,2\pi,\ldots$. Now that $\tilde{\mathbf{v}}_{1}$ is known, Eq.\eqref{eq:continuity_first} is most easily solved by introducing the displacement vector $\tilde{\pmb{\xi}}_{1}(\mathbf{r},t_{0},t_{2},\ldots)$ defined by
\begin{equation}
\tilde{\mathbf{v}}_{1}=\frac{\partial\tilde{\pmb{\xi}}_{1}}{\partial t_{0}};\qquad\tilde{\pmb{\xi}}_{1}(\mathbf{r},t_{0}=0,t_{2},\ldots)=\mathbf{0}\label{eq:xi_definition}
\end{equation}
Substituting for $\tilde{\mathbf{v}}_{1}$ with $\tilde{\pmb{\xi}}_{1}$ in Eq.\eqref{eq:continuity_first}, we find
\begin{equation}
\frac{\partial}{\partial t_{0}}\left[\tilde{n_{1}}+\nabla\cdot\left(\bar{n_{0}}\tilde{\pmb{\xi}}_{1}\right)\right]=0
\label{eq:n_tilde_first}
\end{equation}
which is easily integrated as
\begin{equation}
\tilde{n_{1}}=\tilde{N_{1}}(\mathbf{r})-\tilde{\pmb{\xi}}_{1}\cdot\nabla\bar{n_{0}}-\bar{n_{0}}\nabla\cdot\tilde{\pmb{\xi}}_{1}\label{eq:n_1st_solved}
\end{equation}
where $\tilde{N_{1}}$ is the initial beam density. Eq.\eqref{eq:xi_definition} is integrated as easily using Eq.\eqref{eq:v_betatron_solved}, and we have
\begin{equation}
\tilde{\pmb{\xi}}_{1}=\mathbf{u}_{1}\mbox{sin}t_{0}+\mathbf{u}_{1}\times\mathbf{e}_{z}\left(\mbox{cos}t_{0}-1\right)=\left(\tilde{\mathbf{v}}_{1}-\mathbf{u}_{1}\right)\times\mathbf{e}_{z}
\label{eq:xi_solved}
\end{equation}

These results to first order in $\delta$ can now be used in Eq.\eqref{eq:momentum_second_averaged} and Eq.\eqref{eq:continuity_second_averaged}. After a slightly lengthy calculation, we find
\begin{equation}
<\tilde{n_{1}}\tilde{\mathbf{v}}_{1}>+\bar{n_{0}}<\tilde{\mathbf{v}}_{1}\cdot\nabla\tilde{\mathbf{v}}_{1}>\times\mathbf{e}_{z}=\frac{1}{2}\nabla\times\left(\bar{n_{0}}u_{1}^{2}\mathbf{e}_{z}\right)
\label{eq:betatron_first_curl}
\end{equation}
so that the equation for the evolution of the beam density on the space-charge time scale is given by
\begin{equation}
\frac{\partial\bar{n_{0}}}{\partial t_{2}}+\nabla\cdot\left(\bar{n_{0}}\nabla\bar{\phi_{0}}\times\mathbf{e}_{z}\right)+\alpha^{2}\nabla\cdot\left(\nabla\bar{p_{0}}\times\mathbf{e}_{z}\right)=0\label{eq:density_eq_last_step}
\end{equation}
From the identity $\nabla\bar{p_{0}}\times\mathbf{e}_{z}=\nabla\times(\bar{p_{0}}\mathbf{e}_{z})$ we see that the pressure term vanishes in Eq.\eqref{eq:density_eq_last_step} and we finally have
\begin{equation}
\frac{\partial\bar{n_{0}}}{\partial t_{2}}+\nabla\bar{\phi_{0}}\times\mathbf{e}_{z}\cdot\nabla\bar{n_{0}}=0\label{eq:density_eq_final}
\end{equation}
Eq.\eqref{eq:density_eq_final}, combined with Poisson's equation
\begin{equation}
\nabla^{2}\bar{\phi_{0}}=-\bar{n_{0}}
\label{eq:Poisson_zeroth_again}
\end{equation}
forms the desired closed set of coupled equations, describing the evolution of the beam density under the effect of space-charge forces. It is interesting to note that temperature effects do not enter in this equation when the pressure tensor is gyrotropic, as it is with the ordering we chose for the amplitude of the betatron oscillations. This is the reason why we do not need to add an equation for the evolution of the pressure to the fluid equations \eqref{eq:continuity_moving}-\eqref{eq:Poisson}.

Eq.\eqref{eq:density_eq_final} can be easily interpreted as the advection of the density profile in the velocity field $\mathbf{E}\times\mathbf{B}/B^{2}$, the so-called $\mathbf{E}\times\mathbf{B}$ velocity (there is no - sign in front of $\nabla\bar{\phi_{0}}$ because the Lorentz force is in the opposite direction in the moving frame). This is in agreement with Eq.(1) \cite{Gordon}, and can be explained as follows. In the absence of accelerating gaps, the effect of the betatron oscillations on the density profile averages out to zero on the slow time scale. Thus, the density profile simply follows the slow, averaged motion of the ions. In a homogeneous magnetic field, in the presence of an electric field created by the charges in the bunch themselves, this slow motion is just the $\mathbf{E}\times\mathbf{B}$ motion.

\begin{figure}[ht]
\centering
\includegraphics[width=\textwidth]{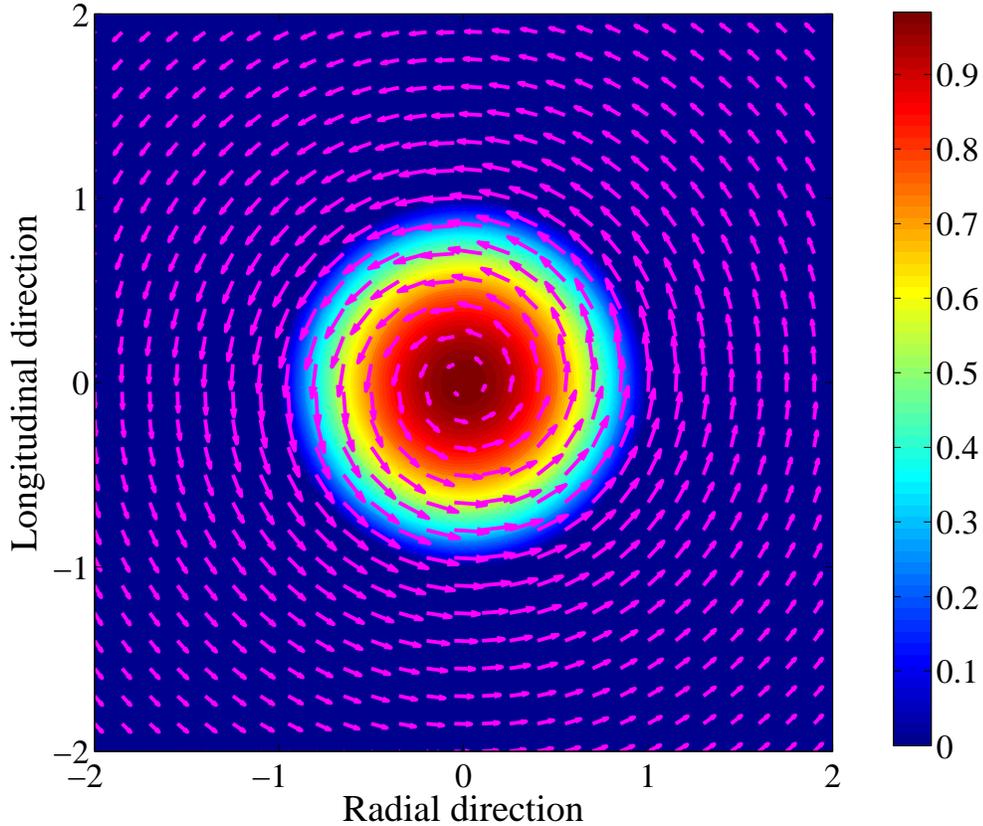}
\caption{Velocity field $\nabla\bar{\phi_{0}}\times\mathbf{e}_{z}$ (magenta arrows) for a cylindrically symmetric density distribution. To compare with figures in Ref.\cite{YangAdel} and Ref.\cite{Adam2}, the beam's transport direction is along the negative $y$ direction.}
\label{fig:vortex_circular}
\end{figure}

\begin{figure}[ht]
\centering
\includegraphics[width=\textwidth]{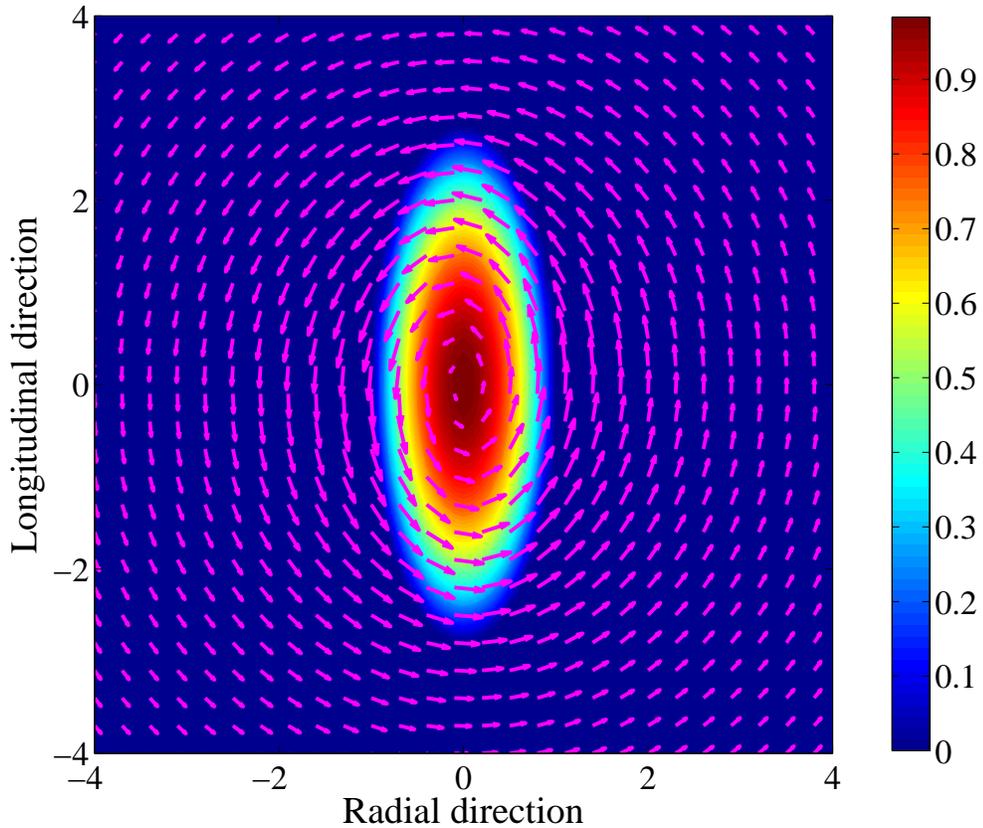}
\caption{Velocity field $\nabla\bar{\phi_{0}}\times\mathbf{e}_{z}$ (magenta arrows) for an elongated density distribution. To compare with figures in Ref.\cite{YangAdel} and Ref.\cite{Adam2}, the beam's transport direction is along the negative $y$ direction.}
\label{fig:vortex_elongated}
\end{figure}

With this intuitive interpretation, it is easy to predict the early stages of the evolution of the beam for the density profiles discussed in the Introduction. The two situations of interest are illustrated in Figures \ref{fig:vortex_circular} and  \ref{fig:vortex_elongated}, which show the $\nabla\bar{\phi_{0}}\times\mathbf{e}_{z}$ velocity field due to a cylindrically symmetric density distribution and an elongated density distribution respectively. If we start, as in Figure \ref{fig:vortex_circular}, with a density profile which is cylindrically symmetric, i.e. such that the beam density is only a function of the distance from the center of the bunch ($\bar{n_{0}}=\bar{n_{0}}(r)$), then the initial electrostatic potential due to the charges in the beam will also be cylindrically symmetric: $\bar{\phi_{0}}=\bar{\phi_{0}}(r)$. This implies that the initial electric field will be along the $\mathbf{e}_{r}$ direction, $\mathbf{E}=E_{r}\mathbf{e}_{r}$, and the convective velocity $\nabla\bar{\phi_{0}}\times\mathbf{e}_{z}$ will be along the $\mathbf{e}_{\theta}$ direction. We thus expect the density profile to rotate around the center of the bunch. Since this density profile is cylindrically symmetric, a rotation of the profile around the center will leave the profile invariant. In other words, the space charge forces in a cylindrically symmetric bunch in a purely drifting region are such that the bunch properties are kept constant. This result agrees with a similar result obtained by Kleeven et al. \cite{Kleeven} with a different method and with a very different set of assumptions.

If, however, the initial density distribution is elongated in the direction of propagation of the beam as in Figure \ref{fig:vortex_elongated}, the cylindrical symmetry will be broken, and the $\mathbf{E}\times\mathbf{B}$ velocity field will distort the beam. Since the electric field is strongest at the extremities of the beam, the distortion will be strongest in these places. This initial situation explains the formation of the spiral arms in the early stages of the beam evolution. In addition, since the electrostatic potential far away from the beam still retains cylindrical symmetry (far away, the beam is seen as a point charge), we expect the $\mathbf{E}\times\mathbf{B}$ convection to act in such a way as to make the beam conform with this cylindrical symmetry. This is the reason for the formation of the often mentioned galaxy-like shape, and after more turns, the formation of a compact, stable cylindrically symmetric core. In the next section, we solve the system of coupled equations Eq.\eqref{eq:density_eq_final} and Eq.\eqref{eq:Poisson_zeroth_again} numerically and confirm these predictions. 

\section{Numerical results}\label{sec:four}
In order to solve Eq.\eqref{eq:density_eq_final} and Eq.\eqref{eq:Poisson_zeroth_again} numerically, we first rewrite them as follows
\begin{equation}
\frac{\partial\bar{n}_{0}}{\partial t}+\delta^{2}\left(E_{x}\frac{\partial\bar{n}_{0}}{\partial y}-E_{y}\frac{\partial \bar{n}_{0}}{\partial x}\right)=0
\label{eq:advection_for_num}
\end{equation}
\begin{equation}
\nabla^{2}E_{x}=\frac{\partial\bar{n}_{0}}{\partial x}\;,\qquad\nabla^{2}E_{y}=\frac{\partial\bar{n}_{0}}{\partial y}
\label{eq:Poisson_elec}
\end{equation}
In Equation \eqref{eq:advection_for_num}, $t$ represents the numerical time steps. Since they are defined in terms of $\omega_{c}$ in the simulations, i.e. in terms of the fast time scale, $\delta^{2}$ appears in the advection term. $E_{x}$ is the electric field in the radial direction, satisfying $E_{x}=-\partial\bar{\phi}_{0}/\partial x$ and $E_{y}$ is the electric field in the longitudinal direction, satisfying $E_{y}=-\partial\bar{\phi}_{0}/\partial y$. We assume open boundary conditions for the electric field, so that $\mathbf{E}$ is simply due to the charges in the beam.

We solve Eq.\eqref{eq:advection_for_num} and Eq.\eqref{eq:Poisson_elec} by discretizing the problem on a regular Cartesian grid. The time integration of Eq.\eqref{eq:advection_for_num} is done with a second order backward difference scheme, except for the first time step which is calculated with a second order trapezoidal scheme. At each time step, we compute the partial derivatives of the density at each grid point with second order centered differences, assuming that the beam density is zero outside the boundary of the computational domain, which is always chosen to be larger than the maximum beam size. We also compute $E_{x}$ and $E_{y}$ at each time step on this grid by numerically solving the two Poisson's equations in Eq.\eqref{eq:Poisson_elec}. Specifically, we calculate $E_{x}$ and $E_{y}$ at interior points by inverting the standard finite difference second order Laplacian matrix, and use the two-dimensional free space Green's function for Poisson's equation to calculate the electric field at the computational boundary:
\begin{equation}
E_{x}(x_{b},y_{b})=\frac{1}{2\pi}\iint dx'dy'\frac{(x_{b}-x')\bar{n}_{0}(x',y')}{(x_{b}-x')^{2}+(y_{b}-y')^{2}}
\label{eq:Green_x}
\end{equation}
\begin{equation}
E_{y}(x_{b},y_{b})=\frac{1}{2\pi}\iint dx'dy'\frac{(y_{b}-y')\bar{n}_{0}(x',y')}{(x_{b}-x')^{2}+(y_{b}-y')^{2}}
\label{eq:Green_y}
\end{equation}
In Eq.\eqref{eq:Green_x} and Eq.\eqref{eq:Green_y}, the subscript $b$ stands for points on the boundary of the computational domain. We use Simpson's rule to perform the numerical integration. Combining all these elements, we have a numerical scheme to integrate Eq.\eqref{eq:advection_for_num} and Eq.\eqref{eq:Poisson_elec} that is second order in both time and space.

For the simulations we present in this article, we choose the same beam parameters as the ones used in \cite{YangAdel} for the study of a 1mA, 3 MeV coasting beam in PSI injector II. With these parameters, we find $\delta^{2}\sim0.8$. This value is at the upper limit of the validity of the multiscale analysis used in this work, but will allow us to compare our results with a large body of PIC simulations performed to investigate space charge effects in PSI injector II \cite{YangAdel},\cite{Adam},\cite{Koscielniak}. In agreement with \cite{YangAdel}, we start with an initial charge distribution given by
\begin{equation}
n(x,y)=N_{0}\mbox{exp}\left(-\frac{x^{2}}{2\sigma_{x}^{2}}-\frac{y^{2}}{2\sigma_{y}^{2}}\right)
\end{equation}

\subsection{Stability of a round beam}
\begin{figure}[ht]
\centering
\includegraphics[width=\textwidth]{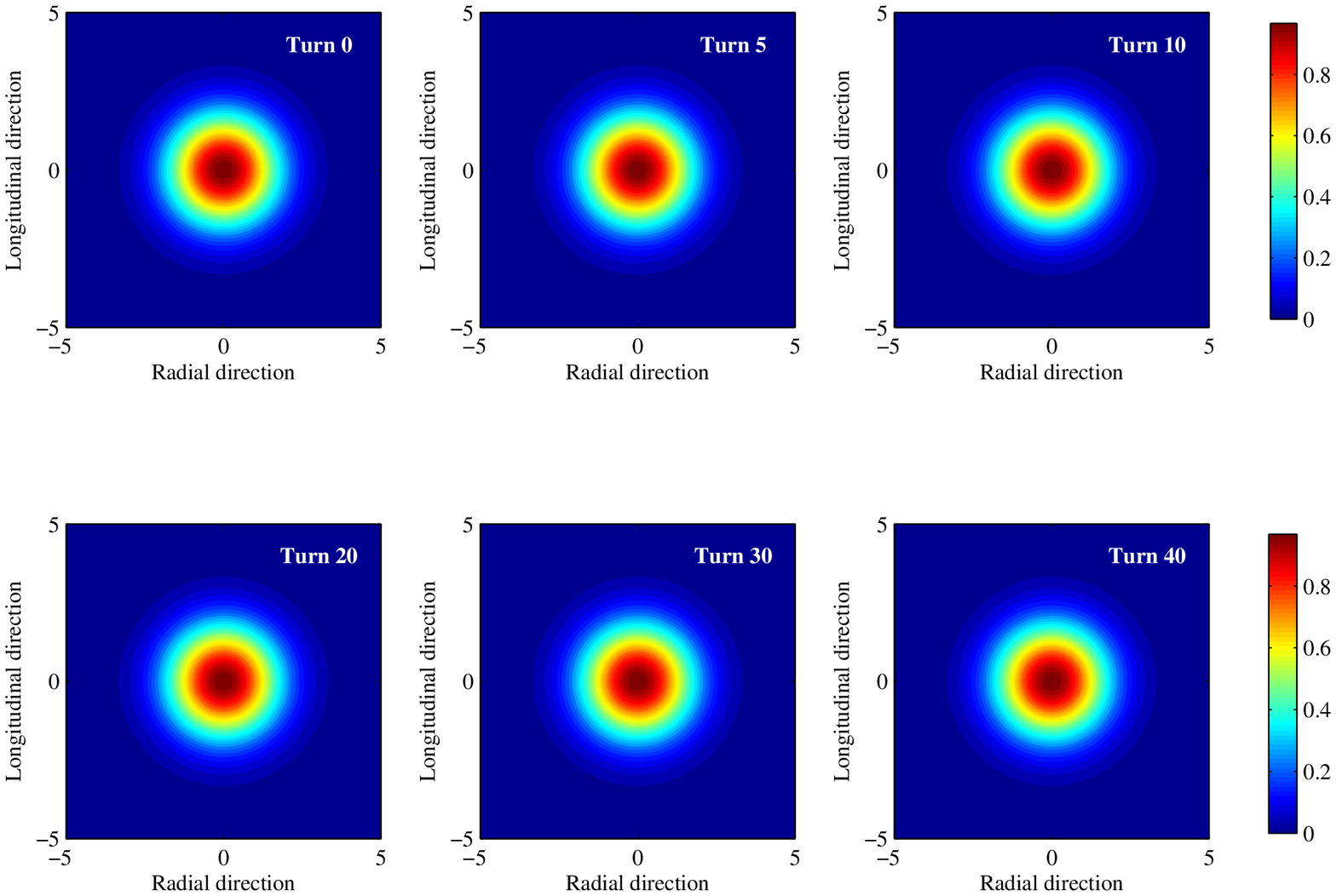}
\caption{Top view of a round coasting beam in the local frame moving with the beam center. Up: turn 0, 5, 10. Down: turn 20, 30, 40. To compare with figures in Ref.\cite{YangAdel} and Ref.\cite{Adam2}, the beam's transport direction is along the negative $y$ direction. $\delta^{2}=0.8$ and $2\sigma_{x}=2\sigma_{y}=2.52\mbox{ mm}$.}
\label{fig:circular_stationary}
\end{figure}

We first demonstrate numerically that a round beam is completely stationary under the effect of space charge forces, as we discussed in the previous section. We choose $2\sigma_{x}=2\sigma_{y}=2.52\mbox{ mm}$ and run our simulation until the beam has completed 40 turns. The results are shown in Figure \ref{fig:circular_stationary}. It is clear in this figure that the charge distribution is not affected by the advection in the $\nabla\phi_{0}\times\mathbf{e}_{z}$ velocity field, and the charge distribution after 40 turns is identical to the initial charge distribution, as expected. In fact, the results shown in Figure \ref{fig:circular_stationary} can be seen as a proof of the accuracy of our numerical method. We can now confidently turn to the study of elongated beams and beam spiraling. 

\subsection{Elongated beams and beam spiraling}
\begin{figure}[ht]
\centering
\includegraphics[width=\textwidth]{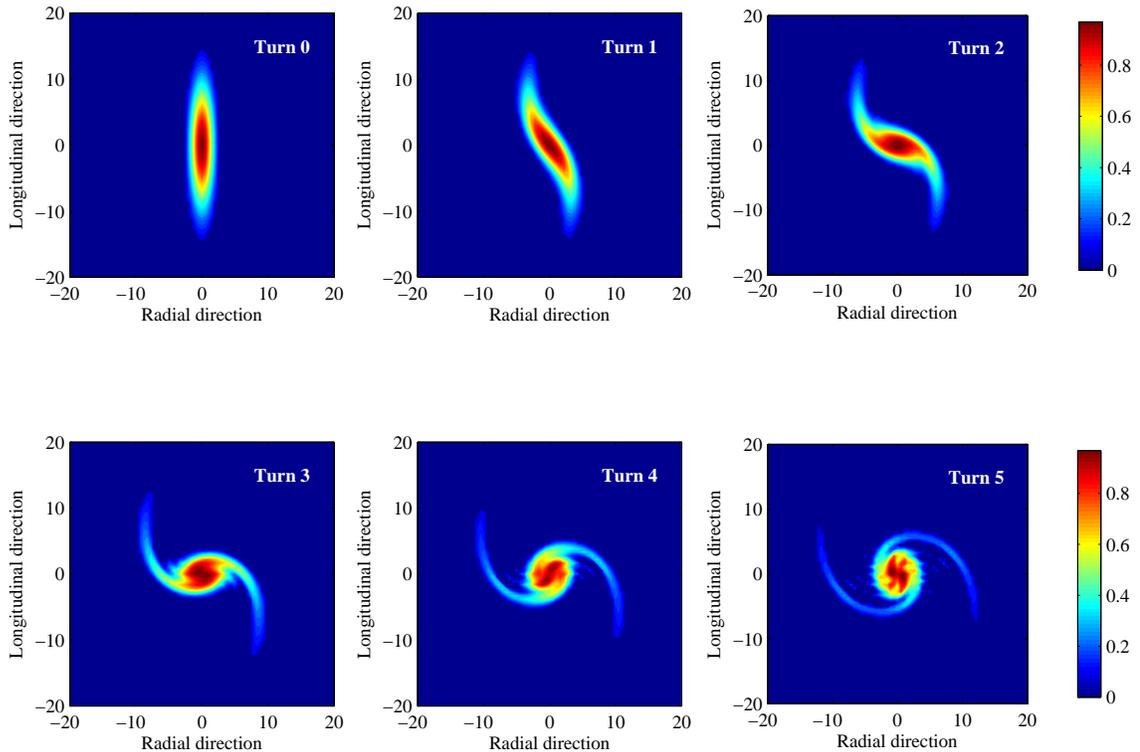}
\caption{Top view of an elongated coasting beam in the local frame moving with the beam center. Up: turn 0, 1, 2. Down: turn 3, 4, 5. To compare with figures in Ref.\cite{YangAdel} and Ref.\cite{Adam2}, the beam's transport direction is along the negative $y$ direction. $\delta^{2}=0.8$, $2\sigma_{x}=2.52\mbox{ mm}$, and $2\sigma_{y}=13.4\mbox{ mm}$}
\label{fig:elongated_adam}
\end{figure}

\begin{figure}[ht]
\centering
\includegraphics[width=\textwidth]{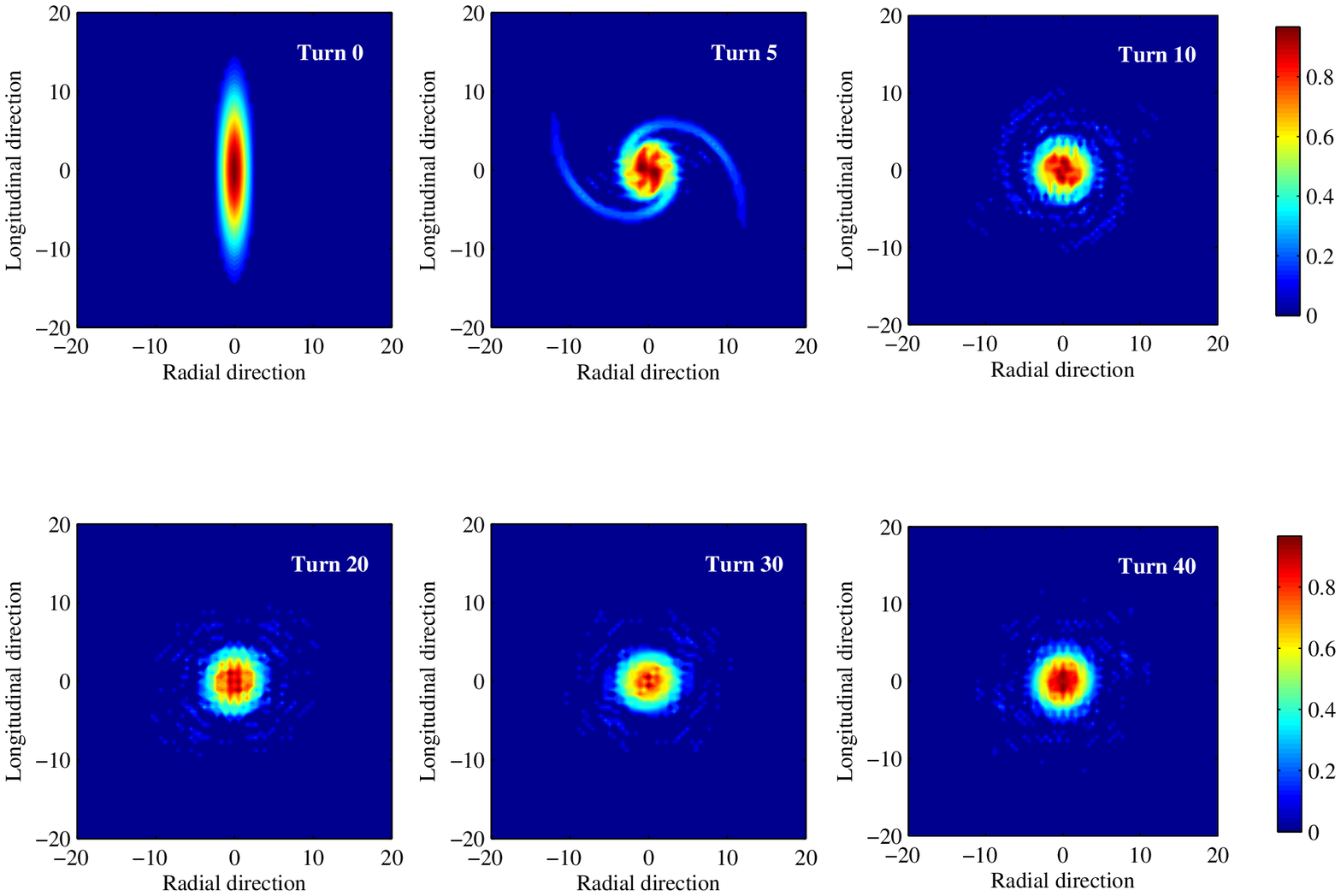}
\caption{Top view of an elongated coasting beam in the local frame moving with the beam center. Up: turn 0, 5, 10. Down: turn 20, 30, 40. To compare with figures in Ref.\cite{YangAdel} and Ref.\cite{Adam2}, the beam's transport direction is along the negative $y$ direction. $\delta^{2}=0.8$, $2\sigma_{x}=2.52\mbox{ mm}$, and $2\sigma_{y}=13.4\mbox{ mm}$}
\label{fig:elongated_adelmann}
\end{figure}

For the study of beam spiraling, we consider an elongated beam with $2\sigma_{x}=2.52\mbox{ mm}$ and $2\sigma_{y}=13.4\mbox{ mm}$ (15$^{\circ}$ phase width) as in \cite{YangAdel}.
The numerical results are shown in Figure \ref{fig:elongated_adam} (early times) and Figure \ref{fig:elongated_adelmann} (later times), where, as in Ref.\cite{YangAdel} and Ref.\cite{Adam2}, we have truncated the beams at 10\% of the maximum charge density. 

By comparing Figure \ref{fig:elongated_adam} with the early stages of the beam evolution shown in Ref.\cite{Adam2}, we observe that there is a strong similarity between our results and those obtained in simulations with the PICS code. We obtain a slightly stronger deformation than seen with PICS in Ref.\cite{Adam2}. This small discrepancy can be attributed to two factors. First, the magnetic geometry is different in the two simulations: the magnetic field in our simulation is uniform and in the $z$ direction, while Adam used an analytic approximation of the magnetic field in the PSI Injector II cyclotron. Second, the shape of the initial charge distribution, which we chose to be identical to that in \cite{YangAdel}, is slightly different. The good agreement between our results and the PICS simulations is not surprising since PICS is a two-dimensional PIC code which relies on the separation between the betatron and space charge time scales to reduce the computational complexity of the PIC simulations. The agreement also suggests that the details of the magnetic geometry do not play a strong role in the radial-longitudinal vortex motion due to space charges.

Comparing Figures \ref{fig:elongated_adam} and \ref{fig:elongated_adelmann} with PICN results \cite{Adam2} and with OPAL-CYCL results \cite{YangAdel}, we conclude that there is qualitative agreement between all descriptions, with the formation of a compact stable core after 40 turns. Furthermore, as in \cite{YangAdel}, we see that a tenuous low density halo persists, even after 40 turns. However, there remains small differences between \cite{YangAdel} and our study. First, the core that is formed after 40 turns is round in our simulations, unlike the slightly elongated core observed in \cite{YangAdel}. Second, the beam spiraling effect appears to be stronger in our results. As before, part of the discrepancy can be explained by the difference in the magnetic geometry assumed in the different simulations. It is also a direct consequence of the two-dimensional description that we have adopted in our study, which leads to a stronger self-electric field, and thus a larger $\nabla\phi_{0}\times\mathbf{e}_{z}$ velocity field. This effect has already been observed in the difference between the PICS and PICN codes \cite{Koscielniak}. Obtaining fully quantitative agreement with OPAL-CYCL on this problem would thus require a three-dimensional fluid treatment, and the inclusion of the details of the PSI injector II magnetic geometry.

\section{Conclusion}\label{sec:five}
In the vast majority of high-intensity cyclotrons and rings, the betatron time scale remains much shorter than the time scale associated with space-charge and temperature effects. This separation of scales can be used advantageously in analytical and numerical studies of space-charge effects on beam transport. In this article we performed a multiscale analysis relying on this scale separation to study the radial-longitudinal vortex motion of an intense charged particle beam in a regime in which the pressure tensor is gyrotropic. As a result of this analysis, we have been able to show that the evolution of the charge distribution on the space-charge time scale is determined by the nonlinear advection equation of the charge distribution in the $\mathbf{E}\times\mathbf{B}$ velocity field, where $\mathbf{E}$ is the self electric field due to space charges and $\mathbf{B}$ is the externally applied magnetic field. The stability of intense round beams is easily understood in this light. A beam with a charge distribution which is only a function of the beam radius has an associated $\mathbf{E}\times\mathbf{B}$ velocity field which is always perpendicular to the beam radius in the radial-longitudinal plane. Such a velocity field cannot modify the charge distribution. Beam spiraling in elongated beams is explained just as easily: in such beams the cylindrical symmetry of the charge distribution is broken, and the $\mathbf{E}\times\mathbf{B}$ velocity field therefore modifies the charge distribution until the cylindrical symmetry is restored. The spiral shape is a natural consequence of the fact that $\mathbf{E}$ is strongest at the extremities of the beam.

Aside from the shape of the beam, the parameter that determines the strength of the spiraling effect is $\delta^{2}=\omega_{p}^{2}/\omega_{c}^{2}$. In cyclotrons and rings with very intense beams and moderate magnetic fields, this parameter is relatively large ($\delta\lesssim 1$) and beam spiraling can play a positive role since a dense round stationary core can be formed after a few turns and the spiral halo can be stripped at low energies. In accelerators with less intense beams or higher fields, however, such that $\delta\ll1$, beam spiraling can have a negative effect since it takes more turns to develop, and high energy beam halos can be formed in this process. 

The good agreement between our numerical simulations and PIC simulations demonstrates that the potential coupling between the radial-longitudinal motion and the vertical motion is not a key physics elements that need to be included to understand space charge effects in the radial-longitudinal plane. This suggests that the simplified fluid equations Eqs.\ref{eq:continuity_moving}-\ref{eq:Poisson}, together with an equation for the evolution of the fluid pressure $p$, might be very well suited for numerical studies of the novel and interesting regime in which $\delta^{2}>1$. In this regime, of interest in some new machines such as the UMER ring \cite{Reiser}, the analytic multiscale expansion used in this work breaks down, but the 2-D fluid equations presented here could still represent a significant reduction in complexity in numerical simulations as compared to PIC simulations. This topic is currently being investigated, with progress to be reported at a later date.
\appendix*
\section{}\label{sec:append}
In this appendix we prove, starting from the collisionless Vlasov equation, that the pressure tensor $\mathbf{P}$ is gyrotropic to lowest order in $\delta$ when the amplitude of the betatron oscillations is small by $\delta$ compared to the size of the beam. 

The collisionless Vlasov equation determining the evolution of the beam distribution function $f(\mathbf{r},\mathbf{v},t)$ is
\begin{equation}\label{eq:Vlasov}
\frac{\partial f}{\partial t}+\mathbf{v}\cdot\nabla f+\frac{e}{m}\left(\mathbf{E}+\mathbf{v}\times\mathbf{B}\right)\cdot\nabla_{\mathbf{v}}f=0
\end{equation}
Let us compare the relative strength of each term in this equation relative to the magnetic term, in the regime where $\mathbf{v}\sim\delta a\omega_{c}$:
\begin{align*}
\frac{\mathbf{E}}{\mathbf{v}\times\mathbf{B}}\sim\frac{ena/\epsilon_{0}}{\delta a \omega_{c}^{2} m/e}\sim\frac{\omega_{p}^{2}}{\delta\omega_{c}^{2}}\sim\delta\qquad\frac{m\mathbf{v}\cdot\nabla f}{e\mathbf{v}\times\mathbf{B}\cdot\nabla_{\mathbf{v}}f}\sim\frac{\delta\omega_{c}}{\omega_{c}}\sim\delta\qquad
\frac{m\partial f/\partial t}{e\mathbf{v}\times\mathbf{B}\cdot\nabla_{\mathbf{v}}f}\sim\frac{\delta\omega_{c}}{\omega_{c}}\sim\delta
\end{align*}
We see that all the terms are small by $\delta$ compared to the magnetic term, so that to lowest order in $\delta$, Eq.\eqref{eq:Vlasov} is:
\begin{equation}\label{eq:Vlasov_ordered}
\mathbf{v}\times\mathbf{B}\cdot\nabla_{\mathbf{v}}f=0+O(\delta)
\end{equation}
Equation \eqref{eq:Vlasov_ordered} implies that to lowest order in $\delta$ the distribution function $f(x,y,z,v_{x},v_{y},v_{z},t)$ has the following particular dependence on $v_{x}$ and $v_{y}$:
\begin{equation}\label{eq:dist_function}
f(x,y,z,v_{x},v_{y},v_{z},t)=f(x,y,z,\sqrt{v_{x}^{2}+v_{y}^{2}},v_{z},t)+O(\delta)
\end{equation}
Thus, to lowest order $f$ is even in $v_{x}$ and $v_{y}$, which implies that to the same order the off-diagonal components of the pressure tensor vanish:
\begin{equation}\label{eq:gyrotropic_append}
\mathbf{P}=p_{\perp}\mathbf{I}+\left(p_{\parallel}-p_{\perp}\right)\mathbf{b}\mathbf{b}+O(\delta)
\end{equation}
with $p_{\perp}=p_{xx}=p_{yy}$ and $p_{\parallel}=p_{zz}$.

\begin{acknowledgments}
The authors would like to thank A. Adelmann from the Paul Scherrer Institut for helpful discussions and for providing the beam parameters allowing comparison between this work and previous work on the beam dynamics in PSI injector II.
\end{acknowledgments}



\begin{thebibliography}{00}
\bibitem{Hockney_Eastwood}R.W. Hockney and J.W. Eastwood, \textit{Computer Simulation Using Particles} (Hilger, New York, 1988)
\bibitem{Cousineau}S. Cousineau, in \textit{Proceedings of the 2003 Particle Accelerator Conference}, (IEEE, Portland, 2003), p.259
\bibitem{YangAdel}J.J. Yang, A. Adelmann, M. Humbel, M. Seidel, and T.J. Zhang, \textit{Physical Review ST Accelerators and Beams} \textbf{13}, 062401 (2010)
\bibitem{Adel}A. Adelmann, P. Arbenz, and Y. Ineichen, \textit{J. Comp. Phys} \textbf{229} (2010) 4554-4566
\bibitem{Holmes}J.A. Holmes, V.V. Danilov, J.D. Galambos, D. Jeon, and D.K. Olsen, \textit{Physical Review ST Accelerators and Beams} \textbf{2}, 114202 (1999)
\bibitem{Zhang}T. Zhang, H. Yao, J. Yang, J. Zhong, S. An, \textit{Nuclear Instruments and Methods in Physics Research A} \textbf{676} (2012) 90-95
\bibitem{Pozdeyev1}E. Pozdeyev, PhD Thesis, MSU, 2003
\bibitem{Adam}S. Adam, \textit{IEEE Trans. Nucl. Sci.} \textbf{32}, 2507 (1985)
\bibitem{Koscielniak}S. Koscielniak and S. Adam, in \textit{Proceedings of the 15th Particle Accelerator Conference} (IEEE, Washington, DC, 1993), p.3639.
\bibitem{Bertrand}P. Bertrand and C. Ricaud, in \textit{Proceedings of the 16th International Conference on Cyclotrons and their Applications} (AIP, East Lansing, Michigan, 2001), p. 379
\bibitem{AdamAdel}S. Adam, A. Adelmann, and R. D\"olling, in \textit{Proceedings of the 16th International Conference on Cyclotrons and their Applications} (AIP, East Lansing, Michigan, 2001), p. 428
\bibitem{Seidel}M. Seidel, S.Adam, A.Adelmann, C.Baumgarten, Y.J.Bi, R.Doelling,H.Fitze, A.Fuchs, M.Humbel, J.Grillenberger, D.Kiselev, A.Mezger, D.Reggiani, M.Schneider, J.J.Yang, H. Zhang, T.J. Zhang, in \textit{Proceedings of IPAC 2010}, Kyoto, Japan, p.1309
\bibitem{Pozdeyev2}E. Pozdeyev, F. Marti, R.C. York, and J. Rodriguez, in \textit{Proceedings of the 2005 Particle Accelerator Conference} (IEEE, Knoxville, Tennessee, 2005), p.159
\bibitem{Bi}Y. Bi, T. Zhang, C. Tang, Y. Huang, and J. Yang, \textit{J. Appl. Phys} \textbf{107}, 063304 (2010)
\bibitem{Chasman}C. Chasman and A.J. Baltz, \textit{Nuclear Instruments and Methods in Physics Research} \textbf{219} (1984) 279-283
\bibitem{Adam2}S. Adam, in Proceedings of the 14th International Conference on Cyclotrons and their Applications, p.446
\bibitem{Davidson}R.C. Davidson, \textit{Physics of Nonneutral Plasmas} (Addison-Wesley, Reading, MA 1990), Chapter 2, p.22
\bibitem{Lund}S.M. Lund and R.C Davidson, \textit{Phys. Plasmas} \textbf{5} 3029 (1998)
\bibitem{Reiser}M. Reiser, P.G. O'Shea, R.A. Kishek, S. Bernal, P. Chin, S. Guharay, Y. Li, M. Venturini, J.G. Wang, V. Yun, W. Zhang, Y. Zou, M. Pruessner, T. Godlove, D. Kehne, P. Haldemann, R. York, D. Lawton, L.G. Vorobiev, I. Haber, and H. Nishimura,  in \textit{Proceedings of the 1999 Particle Accelerator Conference} (IEEE, New York, NY, 1999), p.234
\bibitem{Gordon}M.M. Gordon, in \textit{Proceedings of the 5th International Cyclotron Conference}, Oxford 1969, pp. 305-317
\bibitem{Kleeven} W.J.G.M. Kleeven, H.L. Hagedoorn, and Y.K. Batygin, \textit{Part. Accel.} \textbf{24} (1989), pp.187-210
\end{thebibliography}
\end{document}